\documentclass[preprint,amsmath,superscriptaddress]{revtex4}

\usepackage{graphicx}
\usepackage{epsfig}
\usepackage{hyperref}

\begin{document}
\title{Detection of Rashba spin splitting in 2D organic-inorganic perovskite via precessional carrier spin relaxation}

\author{Seth B. Todd\footnote{Equal contribution}}
\affiliation{Department of Physics and Atmospheric Science,
Dalhousie University, Halifax, Nova Scotia B3H 4R2 Canada}

\author{Drew B. Riley$^{*}$}
\affiliation{Department of Physics and Atmospheric Science,
Dalhousie University, Halifax, Nova Scotia B3H 4R2 Canada}

\author{Ali Binai-Motlagh}
\affiliation{Department of Physics and Atmospheric Science,
Dalhousie University, Halifax, Nova Scotia B3H 4R2 Canada}

\author{Charlotte Clegg}
\affiliation{Department of Physics and Atmospheric Science,
Dalhousie University, Halifax, Nova Scotia B3H 4R2 Canada}

\author{Ajan Ramachandran}
\affiliation{Department of Physics and Atmospheric Science,
Dalhousie University, Halifax, Nova Scotia B3H 4R2 Canada}

\author{Samuel A. March}
\affiliation{Department of Physics and Atmospheric Science,
Dalhousie University, Halifax, Nova Scotia B3H 4R2 Canada}

\author{Ian G. Hill}
\affiliation{Department of Physics and Atmospheric Science,
Dalhousie University, Halifax, Nova Scotia B3H 4R2 Canada}

\author{Constantinos C. Stoumpos}
\affiliation{Department of Chemistry,
Northwestern University, Evanston, Illinois 60208 United States}

\author{Mercouri G. Kanatzidis}
\affiliation{Department of Chemistry,
Northwestern University, Evanston, Illinois 60208 United States}

\author{Zhi-Gang Yu}
\affiliation{Washington State University, Spokane, Washington 99210 United States}

\author{Kimberley C. Hall}
\affiliation{Department of Physics and Atmospheric Science,
Dalhousie University, Halifax, Nova Scotia B3H 4R2 Canada}

\begin{abstract}
{\bf The strong spin-orbit interaction in the organic-inorganic perovskites tied to the incorporation of heavy elements (\textit{e.g.} Pb, I) makes these materials interesting for applications in spintronics.  Due to a lack of inversion symmetry associated with distortions of the metal-halide octahedra, the Rashba effect (used \textit{e.g.} in spin field-effect transistors and spin filters) has been predicted to be much larger in these materials than in traditional III-V semiconductors such as GaAs, supported by the recent observation of a near record Rashba spin splitting in CH$_3$NH$_3$PbBr$_3$ using angle-resolved photoemission spectroscopy (ARPES).  More experimental studies are needed to confirm and quantify the presence of Rashba effects in the organic-inorganic perovskite family of materials.  Here we apply time-resolved circular dichroism techniques to the study of carrier spin dynamics in a 2D perovskite thin film [(BA)$_2$MAPb$_2$I$_7$; BA = CH$_3$(CH$_2$)$_3$NH$_3$, MA = CH$_3$NH$_3$].  Our findings confirm the presence of a Rashba spin splitting via the dominance of precessional spin relaxation induced by the Rashba effective magnetic field.  The size of the Rashba spin splitting in our system was extracted from simulations of the measured spin dynamics incorporating LO-phonon and electron-electron scattering, yielding a value of 10 meV at an electron energy of 50 meV above the band gap, representing a 20 times larger value than in GaAs quantum wells.}
\end{abstract}

\pacs{}

\maketitle
The hybrid organic-inorganic perovskites have gained considerable attention in recent years due to their outstanding performance as absorbing layers in photovoltaics \cite{NREL:web}.  This success has led to a comprehensive research effort with the aim to unravel their photophysical properties \cite{Tanaka:2003, Hirasawa:1994, Miyata:2015, DInnocenzo:2014, YamadaIEEE:2015, Cooke:2015, Fang:2015,Carlos:biexciton, March:exciton, March:manybody, Webber:perov, Manser:review,Sum:review,Herz:2016,Stranks:2013,Xing:2013,Stoumpos:2013}, and to move beyond CH$_3$NH$_3$PbI$_3$ and explore other material compositions including 2D perovskites \cite{Kumar:2014,Marshall:2015,Ke:2017,Saparov:2015,Cao:2015,Stoumpos:2016}.  This burgeoning family of materials offers properties that can be tailored to a broad range of applications in opto-electronics, including photovoltaics \cite{NREL:web,Cao:2015,Smith:2014}, field-effect transistors \cite{Kagan:1999,Chin:2015}, hard radiation detectors \cite{Shibuya:2002}, light-emitting diodes \cite{Ishihara:1989,Tan:2014,Yan:2018}, lasers \cite{Zhu:2015}, and optical sensors \cite{Dou:2014}. 

The hybrid perovskites are characterized by strong spin-orbit coupling (SOC) tied to the constituent heavy elements \cite{Even:rashbareview}.  These strong spin-orbit effects make the perovskite family of materials attractive for applications in semiconductor spintronics and spin opto-electronics \cite{ALS:book,Datta:1990,Hall:2006,Onodera:1994,Rudolph:2003,Hall:2008,Nishikawa:1995,Hall:1999,Giovanni:2015,Odenthal:2017,Giovanni:2016}.  SOC leads to a giant ($\sim$1~eV) splitting of the lowest two conduction bands and influences the band gap and carrier effective masses \cite{Even:2012,Even:2013}.  In conjunction with a lack of inversion symmetry, SOC also leads to an effective magnetic field that lifts the degeneracy of the carrier spin states within each band \cite{ALS:book}.  While this effect has many origins in semiconductors tied to different sources of inversion asymmetry \cite{Lau:2001,Dresselhaus:1955,Rashba:1984,Pfeffer:1999,Hall:2003}, it is most commonly referred to as the Rashba effect after Bychkov and Rashba analyzed the case of structural inversion asymmetry in a two-dimensional electron gas \cite{Rashba:1984}.  In a spintronic device, the effective magnetic field tied to the Rashba effect may be used to operate on the spin state of carriers by inducing precession in a spin field-effect transistor \cite{Datta:1990,Hall:2006}.  The Rashba spin splitting can also enable spin-dependent transport without the need for an external magnetic field or magnetic materials by using a resonant tunnel diode structure \cite{Hallspintrans:2003}.  

Rasbha effects in organic-inorganic perovskites have been the subject of numerous theoretical investigations in recent years \cite{Brivio:2014,Amat:2014,Stroppa:2014,Kim:2014,Kepenekian:2015,Leppert:2016,Yu:2016}.    The predicted Rashba splittings ($\sim$10's of meV) are large compared to the conventional III-V semiconductors upon which most spintronic devices have been based \cite{Even:rashbareview}. Switchable Rashba coupling exploiting the ferroelectric response has also been predicted \cite{Stroppa:2014,Kim:2014}.  Experimental observation of the Rashba effect in the valence band in single crystal CH$_3$NH$_3$PbBr$_3$ was recently demonstrated using ARPES \cite{Niesner:2016}.  The magnitude of the observed Rashba spin splitting (160~meV) was comparable to the largest values predicted theoretically in any of the hybrid perovskite systems \cite{Brivio:2014,Amat:2014,Stroppa:2014,Kim:2014,Kepenekian:2015,Leppert:2016,Yu:2016}.  This result is promising for the prospect of spintronic applications using hybrid perovskites and highlights the need for further experimental studies of Rashba effects in this family of materials. 

Here we report detection of the Rashba splitting in the 2D perovskite (BA)$_2$MAPb$_2$I$_7$ through measurement of carrier spin relaxation using time-resolved circular dichroism techniques.  Measurements of carrier spin dynamics have proven extremely valuable in the study of Rashba coupling in inorganic III-V semiconductor systems \cite{ALS:book}, in which the spin population relaxation time ($T_1$) is sensitive to both the magnitude and wavevector-dependence of the Rashba effective magnetic field \cite{Lau:2001,Hall:2003}.  (BA)$_2$MAPb$_2$I$_7$ exhibits a Ruddlesden-Popper structure with two layers of corner-sharing lead-iodide octahedra separated by long organic cations of butylammonium forming an interdigitating bilayer.  Our experiments indicate that $T_1$ decreases with increasing photo-excitation energy and increases with optically-injected carrier density, both signatures of Rashba precessional spin relaxation \cite{Lau:2001,DP:1972,DK:1986}.  Our simulations of the electron spin dynamics yield a Rashba spin splitting of 10 meV at an electron energy 50 meV above the band gap, a value that is 20 times larger than in GaAs quantum wells \cite{Lau:2001}.  Our findings highlight the utility of spin relaxation studies as a probe for Rashba coupling in the organic-inorganic hybrid perovskite family of materials.     



\section{Rashba effect in perovskite materials} 
\begin{figure}[htb]\vspace{0pt}
    \includegraphics[width=17.0cm]{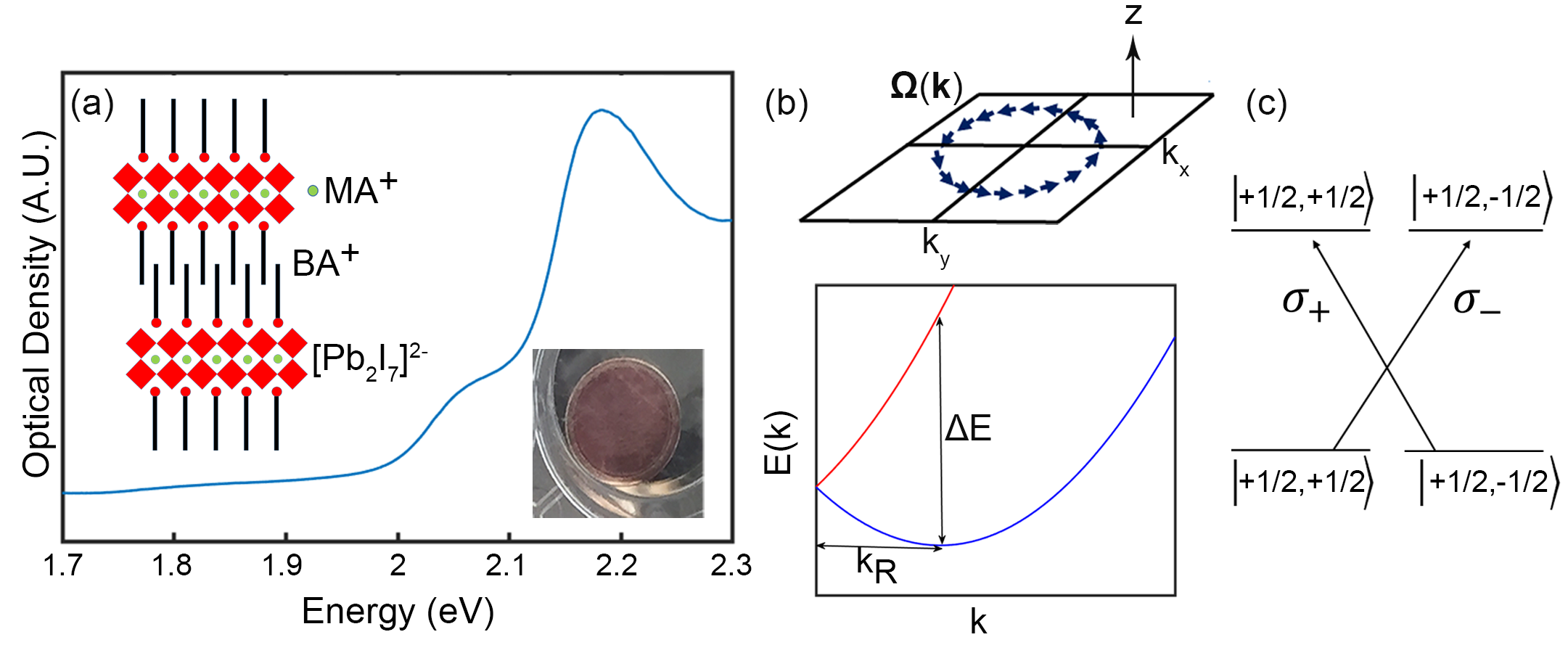}
    \caption{(a) Linear absorption of the sample studied.  Upper inset: (BA)$_2$MAPb$_2$I$_7$ crystal structure; Lower inset: Photo of thin film sample on sapphire substrate.  (b) Upper panel: Rashba effective magnetic field $\mathbf{\Omega}$($\mathbf{k}$) showing the variation of the direction at a fixed value of $\lvert \mathbf{k} \rvert$. Lower panel: Energies of the spin eigenstates (E$_{+}$: red curve, E$_{-}$: blue curve) as a function of in-plane wave vector.  (c) Optical selections rules for interband transitions at the band gap of (BA)$_2$MAPb$_2$I$_7$. }
    \label{figure1}
\end{figure}

In systems with no center of inversion symmetry, the spin-orbit interaction leads to an effective magnetic field $\mathbf{\Omega}(\mathbf{k})$ with a direction that determines an effective spin quantization axis at each value of the electron wavevector $\mathbf{k}$ and a magnitude that determines the energy separation between spin-up and spin-down band states.  The crystal structure will dictate the wavevector dependence of $\mathbf{\Omega}(\mathbf{k})$.  For the organic-inorganic perovskites, symmetry breaking is tied to tilting and associated polar distortions of the metal-halide octahedra, which are influenced by the constituent organic cations.   For the 2D perovskite family (BA)$_2$(MA)$_{n-1}$Pb$_n$I$_{3n+1}$, the tilting of the lead iodide octahedra is determined by the interplay between the long organic BA cations and the smaller MA cations, which induce out-of-plane and in-plane octahedral distortions respectively \cite{Stoumpos:2016}.   For the n = 2 structure studied in this work [Fig.~\ref{figure1}(a)], x-ray diffraction indicates a noncentrosymmetric structure with a polar (C$_{2v}$) space group at room temperature \cite{Stoumpos:2016}.  As a result, a nonzero Rashba spin splitting is expected for this system.  The Rashba Hamiltonian is given by \cite{Rashba:1984,Kepenekian:2015}
\begin{equation}
\label{HRashba}
H_{R} = \frac{\hbar}{2} \mathbf{\Omega} (\mathbf{k}) \cdot \mathbf{\sigma} 
\end{equation}
\noindent
where $\mathbf{k} = (k_x, k_y)$ is the electron wavevector in the x-y plane perpendicular to the stacking direction, $\mathbf{\sigma}$ is the vector of Pauli spin matrices, and
\begin{align*}
        & \Omega_x = \frac{2}{\hbar}\lambda_R k_y&&\Omega_y = - \frac{2}{\hbar}\lambda_R k_x.
\end{align*}
In this case,
\begin{equation}
\label{HRashba_pinwheel}
H_{R}(\mathbf{k})= \lambda_R(k_x\sigma_y-k_y\sigma_x).
\end{equation}
\noindent
and the spin eigenstates are given by
\begin{equation}
\label{rashbabranches}
E_{\pm} (\mathbf{k}) = \frac{\hbar^2 k^2}{2m}\pm\lambda_R \lvert \mathbf{k} \rvert
\end{equation}
\noindent
where $\lambda_R$ is the Rashba coupling parameter. The Rashba effective magnetic field and the dispersion relations for the two spin eigenstates are shown in Fig~\ref{figure1}(b).  



The Rashba effective magnetic field leads to precessional carrier spin relaxation.  This mechanism was first discussed in the context of III-V semiconductors by D'Yakonov and Perel \cite{DP:1972} and later refined for semiconductor quantum wells by D'Yakonov and Kachorovskii \cite{DK:1986}.  Due to the optical selection rules [Fig.~\ref{figure1}(c)], excitation with circularly polarized light leads to a fully spin polarized distribution of electrons and holes.  The carrier spins will be aligned parallel or antiparallel to the light propagation direction (the $\mathbf{z}$ direction in Fig.~\ref{figure1}(b)) depending on the carrier type (electron or hole) and the circular polarization state of the light field.  $\mathbf{\Omega}(\mathbf{k})$ will cause these initially polarized carriers spins to precess.  Since the direction of $\mathbf{\Omega}(\mathbf{k})$ varies with $\mathbf{k}$, the net spin polarization of the ensemble decays.  Our interest here is on the longitudinal spin lifetime in the absence of an external magnetic field ($T_1$).  A characteristic feature of this spin relaxation mechanism is that $T_1$ is proportional to the carrier scattering rate \cite{DK:1986}, referred to as motional narrowing:  The larger the rate of scattering, the slower the spin relaxation because the carrier spins do not have as much time to undergo precession between scattering events.  

The rate of precessional spin relaxation is proportional to the square of the Rashba effective magnetic field \cite{Lau:2001}.  Measurement of $T_1$ therefore provides a sensitive probe of the strength of Rashba coupling in a semiconductor material.  For instance, in GaAs quantum wells the spin lifetime at 300~K is 100~ps \cite{Terauchi:1999} and the corresponding spin splitting has been calculated to be $<$1 meV \cite{Lau:2001}.  In contrast,  in quantum wells formed from the InAs/GaSb/AlSb system possessing a much larger SOC, calculated spin splittings are $\sim$ 10~meV  or larger \cite{Lau:2001,Hallspintrans:2003}, and measured spin lifetimes in these materials under 1 ps have been reported \cite{Hall:1999,Hall:2003}.    

\section{Carrier charge and spin dynamics}
\begin{figure}[htb]\vspace{0pt}
    \includegraphics[width=17.0cm]{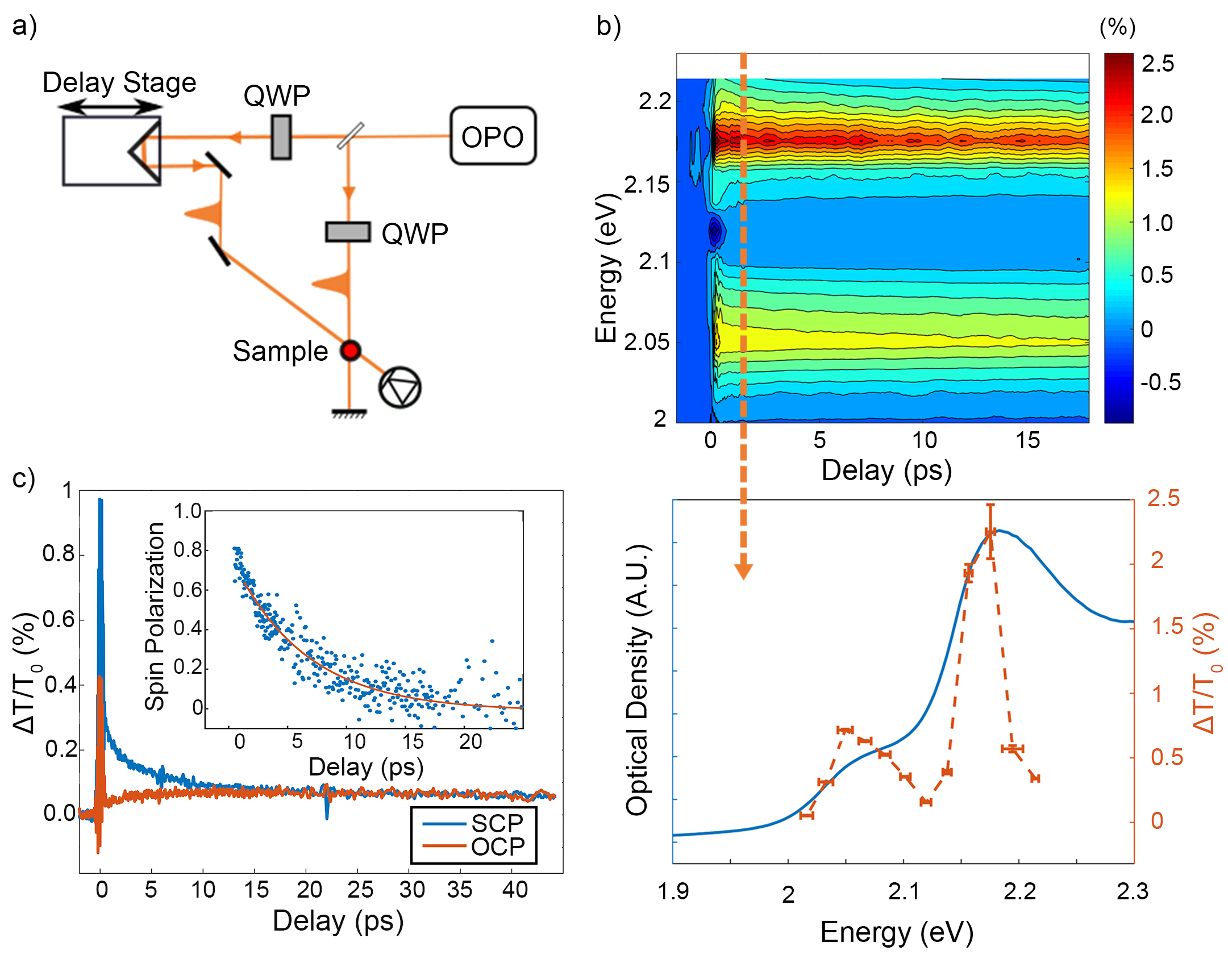}
    \caption{(a) Schematic diagram of the time-resolved circular dichroism experiment.  (b) Upper panel: Differential transmission ($\frac{T - T_0}{T_0}$) as a function of laser tuning and interpulse delay with orthogonal linear polarizations in the pump and probe pulses to probe the spin-independent carrier dynamics. Lower panel: The magnitude of the $\frac{T - T_0}{T_0}$ signal at 2 ps delay, plotted together with the linear absorption spectrum.  For these results, the pump pulse fluence was 0.12 $\mu$J/cm$^{2}$.  The differential transmission signal indicates separate peaks tied to excitons (2.05~eV) and unbound electron-hole pairs above the band gap (2.18~eV). (c) Results of circular dichroism experiments for a laser tuning of 2.15~eV.  The blue (red) curve shows the results for the same (opposite) circular polarization states in the pump and probe beams. Inset: Degree of spin polarization versus delay.  The red curve indicates the results of fitting to a single exponential with a decay time of (10$\pm$1) ps.}
    \label{figure2}
\end{figure}

The results of differential transmission measurements [($\frac{T - T_0}{T_0}$), where $T$ ($T_0$) is the transmission of the probe pulse in the presence (absence) of the pump beam, see Fig.~\ref{figure2}(a)] for linearly-polarized pump and probe pulses is shown in Fig.~\ref{figure2}(b).  The use of linearly-polarized pulses enables measurement of the average state filling signal associated with both spin populations.  The magnitude of the differential transmission signal at $\tau$ = 2 ps is plotted alongside the linear absorption spectrum in the lower panel of Fig.~\ref{figure2}(b).  A Tauc analysis of the linear absorption results (see Supporting Information) indicates a band gap of (2.12$\pm$0.05)~eV.  The peak at 2.05~eV is therefore attributed to the state filling response of excitons, and the peak at higher energies is due to unbound electron-hole pairs on interband transitions above the band gap.  Fitting the results in Fig.~\ref{figure2}(b) for above band gap excitation to a double exponential decay yields a carrier thermalization time of (7$\pm$1) ps, followed by recombination on a time scale $\sim$1~ns.  These spin-independent relaxation kinetics are in line with previous studies in phenylethylammonium lead iodide (PEA)$_2$(MA)$_{n-1}$Pb$_n$I$_{3n+1}$ 2D perovskite \cite{Milot:2016}.    

$T_1$ was measured using time-resolved circular dichroism. A circularly-polarized pump pulse is used to inject a spin-polarized carrier distribution.  When the probe has the same (opposite) circular polarization as the pump, the state filling signal is tied to the majority (minority) spin population.  The convergence of these two state filling signals with increasing interpulse delay indicates carrier spin relaxation. Fig.~\ref{figure2}(c) shows the results of circular dichroism experiments for the same pump pulse fluence as in Fig.~\ref{figure2}(b) under excitation at 2.15~eV (30~meV above the band gap).   A strong oscillatory signal associated with the coherent artifact \cite{note:CA} is superimposed on the state-filling response for delay values within the range of overlap between the pump and probe pulses.   Beyond this overlap region, the differential transmission signals for the same circular polarization (SCP) and opposite circular polarization (OCP) are clearly resolved, indicating a difference in the magnitude of the state-filling response tied to the majority and minority carrier spin populations.  The degree of spin polarization of the carrier distribution may be extracted from the results in Fig.~\ref{figure2}(c) by taking the difference of the SCP and OCP responses and dividing by the sum (see Supporting Information for further details).  The result of this analysis is shown in the inset to Fig.~\ref{figure2}(c).  The decay of the carrier spin polarization fits to a single exponential, yielding $T_1$ = (10$\pm$1)~ps.  Recent calculations indicate similar effective masses for electrons and holes in this system \cite{Wang:2017}.  The resulting similarity in the density of states in the valence and conduction bands has the consequence that the two carrier species contribute to a similar extent to the magnitude of the measured state filling signal.  The observation of a single exponential decay, together with the large value of the initial degree of spin polarization, points to a similar spin lifetime for electrons and holes.

The measured spin lifetimes for a range of laser tunings ($E_{ex}$) and pump pulse fluence values are shown in Fig.~\ref{figure3}(a) and Fig.~\ref{figure3}(b), respectively.   In order to detect evidence of Rashba precessional spin relaxation, the spin lifetime for carriers above the band gap is of interest.  Fig.~\ref{figure3}(a) shows the spin lifetime versus laser detuning ($E_{ex} - E_g$), which together with the carrier masses determines the electron and hole energies relative to the band edges.  The spin lifetime decreases for increasing laser detuning, indicating that spin relaxation is more rapid for carriers with larger kinetic energies.  The spin lifetime was also observed to increase with increasing pump pulse fluence (Fig.~\ref{figure3}(b)).  Increasing the density of optically-injected carriers therefore results in a longer spin lifetime.

\section{Discussion}
\begin{figure}[htb]\vspace{0pt}
    \includegraphics[width=17.0cm]{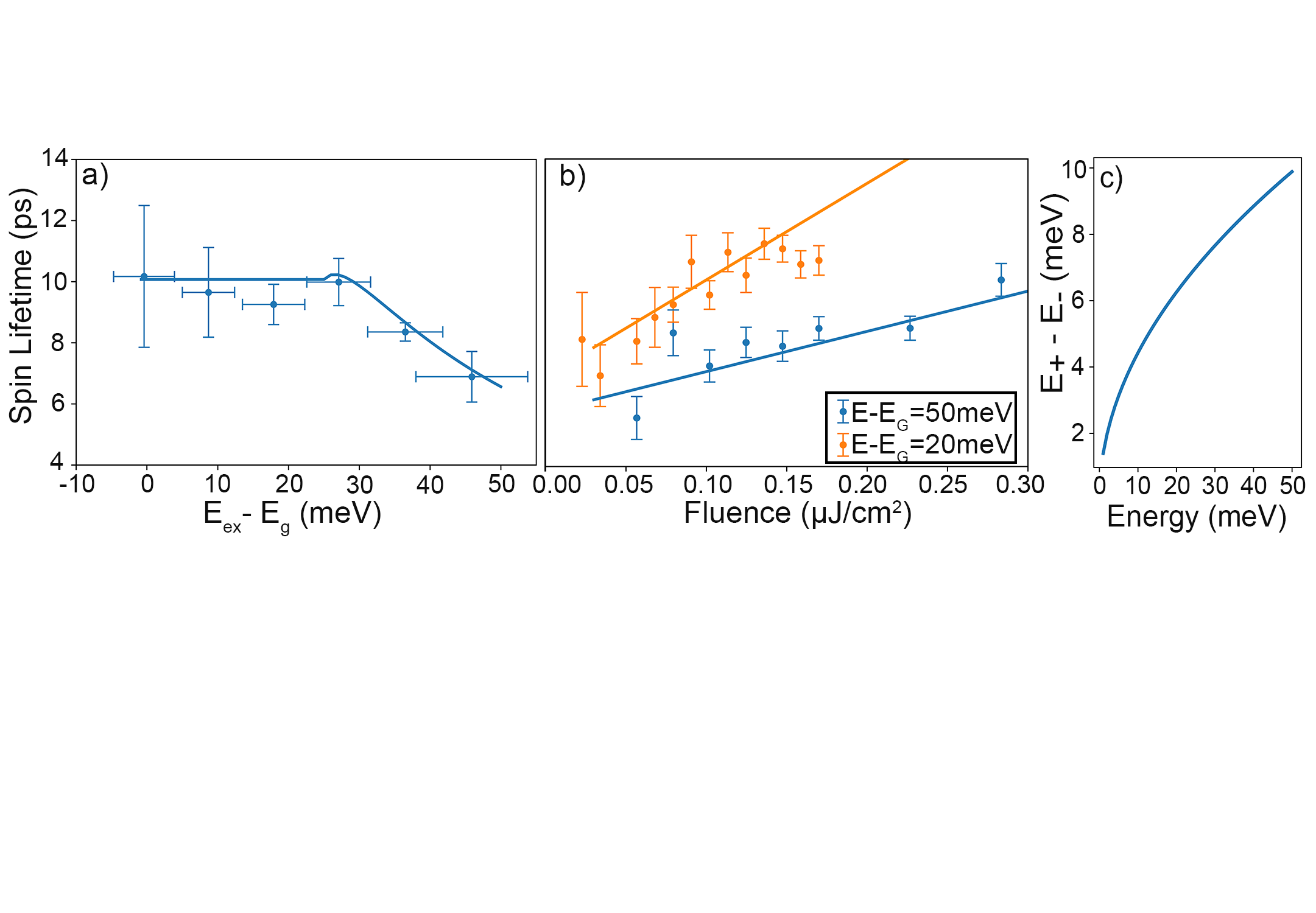}
    \caption{Measured dependence of the spin lifetime on laser detuning [(a)] and laser pulse fluence [(b)]. The solid curves show the results of simulations of the carrier spin dynamics taking the Rashba coupling strength as a fitting parameter.  (c) Calculated conduction band spin splitting using the optimum value of the Rashba coupling strength determined from the numerical simulations, indicating a spin splitting of 10~meV at an energy of 50~meV above the band edge.}
    \label{figure3}
\end{figure}

The trends observed in Fig.~\ref{figure3}(a) and Fig.~\ref{figure3}(b) are consistent with precessional spin relaxation induced by the Rashba effective magnetic field.  Since $\mathbf{\Omega} (\mathbf{k})$ has a magnitude and direction that varies with wavevector, precession causes the initially spin-polarized distribution of carriers to decay.  Scattering of carriers with phonons, impurities, and each other will slow this spin relaxation process by inducing rapid changes in the value of $\mathbf{\Omega} (\mathbf{k})$, resulting in a smaller degree of precession between scattering events. This motional narrowing feature is clearly reflected in the observation of an increase in $T_1$ with increasing carrier density, indicating slower spin relaxation with an increased rate of carrier-carrier scattering.  The observation of a decrease in $T_1$ with increasing excess energy is also consistent with Rashba precessional spin relaxation.  Since the carrier thermalization time was found to be similar to the spin lifetime, the carriers remain hot during spin relaxation, with a temperature determined by the laser detuning.  A higher average carrier kinetic energy will lead to the occupation of states at larger $\mathbf{k}$ values.  Since the magnitude of the effective magnetic field increases with $\lvert \mathbf{k} \rvert$, hot carriers will undergo more rapid spin relaxation.  

We note that the results in Fig.~\ref{figure3}(b) are opposite to the trend expected for the Elliot Yafet spin relaxation process \cite{Elliot:1954Yafet:1963}, for which scattering events themselves lead to spin decay rather than precession between scattering events.  The spin relaxation process tied spin-flip scattering between electrons and holes (the Bir Aronov Pikus mechanism) \cite{BAP:1976}, is expected to be weak in perovskite materials due to the small value of the exciton exchange splitting \cite{Odenthal:2017}, a feature that has been attributed to the spatial separation of the electron and hole wave functions within the unit cell.

An estimate of the magnitude of the Rashba spin splitting in the 2D perovskite structure studied here was obtained from simulations of the measured carrier spin dynamics.  These simulations incorporated polar optical phonon scattering, which is expected to be stronger than acoustic phonon scattering at 300~K \cite{Ni:2017}, as well as 2D electron-electron scattering mediated by the Coulomb interaction in the nondegenerate limit.  For these calculations, the carrier effective masses and dielectric constant were taken from recent first principle calculations for (BA)$_2$PbI$_4$ \cite{Wang:2017}, and the LO phonon energy was taken from Ref.~\onlinecite{Ni:2017} (see Supporting Information).  $\lambda_R$ was varied to obtain the best fit to the experimental results.  The resulting fits are shown as the solid curves in Fig.~\ref{figure3}(a) and Fig.~\ref{figure3}(b).  The Rashba spin splitting extracted from the fitting process is shown in Fig.~\ref{figure3}(c).  A splitting of 10 meV is found at an energy of 50 meV above the conduction band minimum.  For comparison, for GaAs quantum wells at 300~K, a much smaller spin splitting of 0.5 meV at this energy was found from the analysis of spin lifetime results \cite{Lau:2001}.  The spin-independent pump probe experiments in Fig.~\ref{figure2}(b) indicated a thermalization time of 7 ps, resulting in an effective carrier temperature somewhat cooler than determined by the laser detuning alone.  Our simulations, which neglect thermalization effects, therefore provide a lower bound to the Rashba spin splitting in this material.  

Our observation of a much larger Rashba spin splitting in (BA)$_2$MAPb$_2$I$_7$ than in inorganic III-V quantum wells is expected from the strong spin-orbit interaction in the organic-inorganic family of materials \cite{Even:rashbareview}. A Rashba effect was observed experimentally for the first time in an organic-inorganic perovskite (in the valence band of CH$_3$NH$_3$PbBr$_3$) using ARPES \cite{Niesner:2016}, yielding an extremely large spin splitting of 160~meV.  A 40~meV Rashba splitting was also recently found using electro-absorption techniques and DFT calculations in the 2D perovskite (PEA)$_2$PbI$_4$ \cite{Zhai:2017}.  These large Rashba spin splittings are promising for the prospect of spintronic devices based on the organic-inorganic perovskites.

\section{Conclusions}
We have measured the spin relaxation time for optically-injected carriers in thin films of the 2D perovskite (BA)$_2$MAPb$_2$I$_7$ using time-resolved circular-dichroism techniques.  Our results indicate a value for $T_1$ of 10 ps at 300~K.  The measured dependence of the spin lifetime on the laser excess energy and excited carrier density indicate precessional spin relaxation caused by the Rashba effect.  Simulations of the measured carrier spin dynamics incorporating LO phonon and electron-electron scattering enabled the extraction of a Rashba spin splitting of 10 meV at an electron energy of 50 meV above the band gap.   Our experiments, which confirm the presence of a Rashba spin splitting experimentally in a 2D perovskite material, open the door to studies of spin relaxation kinetics as a probe of Rashba effects in a wide range of 2D and 3D perovskite materials.  For instance, the extension to larger thicknesses of the inorganic layer (\textit{e.g.} (BA)$_2$(MA)$_{n-1}$Pb$_{n}$I$_{3n+1}$ with n $>$ 2) would enable an interesting comparison to the quantum well-width dependence in III-V semiconductors \cite{Tackeuchi:1997}.  Our findings shed-light on the spin-related properties of the organic-inorganic perovskites and point to the potential for spintronic devices based on these materials.



\clearpage
\clearpage
\clearpage
\clearpage

\section{Acknowledgements}     
Work at Dalhousie University was supported by the Natural Sciences and Engineering Research Council
of Canada.  Work at Northwestern University was supported by grant SC0012541 from the US Department of Energy, Office of Science.  Work at Washington State University was supported by grant W911NF-17-1-0511 from the US Army Research Office.







\begin{references}
\bibitem{note:CA}  The coherent artifact is caused by four-wave mixing on the probed optical transitions, and is only present for temporally overlapped pump and probe pulses.
\end{references}
\end{document}